\documentclass[english,amssymb,floatfix]{revtex4}
\usepackage[T1]{fontenc}
\usepackage[latin9]{inputenc}
\usepackage{amsmath}
\usepackage{amssymb}
\usepackage{graphicx}
\usepackage{esint}

\makeatletter

\@ifundefined{textcolor}{}
{%
 \definecolor{BLACK}{gray}{0}
 \definecolor{WHITE}{gray}{1}
 \definecolor{RED}{rgb}{1,0,0}
 \definecolor{GREEN}{rgb}{0,1,0}
 \definecolor{BLUE}{rgb}{0,0,1}
 \definecolor{CYAN}{cmyk}{1,0,0,0}
 \definecolor{MAGENTA}{cmyk}{0,1,0,0}
 \definecolor{YELLOW}{cmyk}{0,0,1,0}
}

\usepackage{amsfonts}\usepackage{bbm}\usepackage{epsfig}\usepackage{graphics}
\usepackage{ifpdf}\ifpdf
\DeclareGraphicsExtensions{.pdf,.png}\usepackage{pgf}\usepackage{tikz}\usepackage{epstopdf}

\else
\DeclareGraphicsExtensions{.eps,.png}
\fi

\graphicspath{{./}}

\newcommand{\lsim}{\,\lower2truept\hbox{${<\atop\hbox{\raise4truept\hbox{$\sim$}}}$}\,}\newcommand{\gsim}{\,\lower2truept\hbox{${>\atop\hbox{\raise4truept\hbox{$\sim$}}}$}\,}\newcommand{\pp}{~~~.}

\newcommand{\be}{\begin{equation}}\newcommand{\ee}{\end{equation}}\newcommand{\bea}{\begin{eqnarray}}\newcommand{\eea}{\end{eqnarray}}\newcommand{\beann}{\begin{eqnarray*}}\newcommand{\eeann}{\end{eqnarray*}}

\newcommand{\eprint}[1]{\url{arXiv:#1}}

\def\Planck{\textit{Planck}}

\makeatother

\usepackage{babel}

\makeatother

\usepackage{babel}

\makeatother

\usepackage{babel}

\makeatother

\usepackage{babel}

\makeatother

\usepackage{babel}

\begin{document}

\title[Testing Dark Energy and Modified Gravity with \Planck]
{Testing modified gravity with \Planck: the case of coupled dark energy}

\author{Valeria Pettorino$^{1}$}
\affiliation{
$^1$ D\'epartement de Physique Th\'eorique and Center for Astroparticle Physics, Universit\'e de Gen\`eve, 24 quai Ernest Ansermet, CH--1211 Gen\`eve 4, Switzerland}

\begin{abstract}
The \Planck\ collaboration has recently published maps of the Cosmic Microwave Background (CMB) radiation, in good agreement with a $\Lambda$CDM model, a fit especially valid for multipoles $\ell> 40$. We explore here the possibility that dark energy is dynamical and gravitational attraction between dark matter particles is effectively different from the standard one in General Relativity: this is the case of coupled dark energy models, where dark matter particles feel the presence of a fifth force, larger than gravity by a factor $\beta^2$. We investigate constraints on the strength of the coupling $\beta$ in view of \Planck\ data. 
Interestingly, we show that a non-zero coupling is compatible with data and find a likelihood peak at $\beta = 0.036 \pm 0.016$  (\Planck\ + WP + BAO) (compatible with zero at 2$\sigma$). The significance of the peak increases to $\beta = 0.066 \pm 0.018$  (\Planck\ + WP + HST) (around 3.6$\sigma$) when \Planck\ is combined to Hubble Space Telescope data.
This peak comes mostly from the small difference between the Hubble parameter determined with CMB measurements and the one coming from astrophysics measurements. In this sense, future observations and further tests of current observations are needed to determine whether the discrepancy is due to systematics in any of the datasets. Our aim here is not to claim new physics but rather to show how Planck data can be used to provide information on dynamical dark energy and modified gravity, allowing us to test the strength of an effective fifth force between dark matter particles with precision smaller than 2$\%$.
\end{abstract}

\date{\today}
\maketitle

\section{Introduction}
Our knowledge of the Cosmic Microwave Background (CMB) has impressively grown in the last few months. 
The South Pole Telescope (SPT, \cite{k11}) and Atacama Cosmology Telescope (ACT, \cite{act_2011}) allowed to detect the first compelling evidence of CMB lensing, pushing our knowledge of the temperature power spectrum of primordial acoustic oscillations up to multipoles $l \sim 3000$ and very small scales. More recently, the \Planck\ collaboration has released the first cosmological papers providing the highest resolution, full sky, maps of the CMB temperature anisotropies, with an accuracy now set by fundamental astrophysical limits. The corresponding analysis of cosmological parameters has been illustrated in \cite{Planck_params}. This extends and increases the resolution of previous measurements of temperature power spectrum (Wilkinson Microwave Anisotropy Probe 9, \cite{wmap9}).

 \Planck\ data are in good agreement with a $\Lambda$CDM cosmology, especially for $\ell > 40$. They can provide interesting bounds on the Early Universe, putting stringent limits to primordial non-Gaussianity \cite{Planck_nonG} and testing inflationary models \cite{Planck_inflation}. They can be used to test isotropy \cite{Planck_isotropy} and infer properties of large scale structures via the $SZ$ effect \cite{Planck_SZ}. In general, they can be used to estimate cosmological parameters \cite{Planck_params} assuming a given model for the background and evolution of perturbations as well as for the foreground components. Such a detailed picture of primordial fluctuations is also able to provide constraints on late time cosmology, for example via CMB lensing \cite{Planck_lensing, acquaviva_baccigalupi_2006, 2013arXiv1305.0829C}. First tests of late time cosmology using \Planck\ data have been presented in \cite{Planck_params} on simple parametrizations of the equation of state and Early Dark Energy. Here we want to show further how CMB probes such as \Planck\ are powerful tests also for dynamical Dark Energy and extensions of General Relativity that modify gravitational interactions, extending and updating the work done in \cite{pettorino_etal_2012}. 

The simplest framework for dark energy models considers dark energy as a
cosmological constant $\Lambda$, contributing to about $68\%$ of the total
energy density in the universe and providing late time cosmic acceleration, 
while Cold Dark Matter (CDM) represents about $27\%$ ($\Lambda$CDM model).  Though theoretically in good agreement with present
observations, a cosmological constant is somewhat unpleasantly
affected by coincidence and fine-tuning problems which seem unavoidable in
such a framework. 
In a $\Lambda$CDM cosmology, Dark Energy density $\rho_{\Lambda}$ is constant; however we usually describe constituents of the Universe in terms of ratios of densities $\Omega_i = \rho_i / \rho_{cr}$, where the subscript indicates the (i) constituent of the Universe (DE, CDM, radiation) and $\rho_{cr}$ is the energy density corresponding to a spatially flat geometry. In particular, in a $\Lambda$CDM, $\Omega_{\Lambda}$ is completely negligible in the past and changes rapidly just at recent times, increasing from nearly zero to about $68 \%$ of the total energy budget. In this framework, for the whole evolution of the Universe, its equation of state is $w = -1$.

Given the degeneracy in the reionization epoch, WMAP polarization (WP) likelihood \cite{2012arXiv1212.5225B, wmap9} can be used in addition to \Planck\ likelihood \cite{Planck_params}. The combination with astrophysical probes further tightens bounds on the equation of state $w$. Such external measurements include geometrical measurements like Baryonic Acoustic Oscillations (BAO) \cite{2005MNRAS.362..505C, 2005ApJ...633..560E}, which are in nice agreement with \Planck\ results for a $\Lambda$CDM model, as well as constraints on the Hubble parameter or Supernovae (see \cite{Planck_params} for a detailed discussion on the different datasets). In particular, when a constant $w$ is assumed for dark energy, this parameter is constrained to be $w = -1.13^{+0.24}_{-0.25} $ at 95$\%$ C.L. when using \Planck\ + WP+ BAO \cite{Planck_params}, in good agreement with $w = -1$; when measurements on $H_0$ from the Hubble Space Telescope (HST) \cite{riess_etal_2011} are combined with \Planck\ + WP, \cite{Planck_params} found $w = -1.24^{+0.18}_{-0.19}$ at 95$\%$ C.L. which is in tension with $w = -1$ at more than 2 $\sigma$ level. Such discrepancy, however, has to be treated with care as it may very well depend on systematics in the measurement of $H_0$.

Many alternative models have been proposed, though it is fair to say
that so far none of them  completely avoids the fine-tuning and coincidence problems nor provides a better fit to data than $\Lambda$CDM. Some
encouraging arguments have been put forward in the framework of dynamical dark
energy models, where a scalar field (quintessence or cosmon) rolls down a
 suitable potential \cite{wetterich_1988, ratra_peebles_1988}.
Small changes of the equation of state around its present value $w_0$ have also been tested, using the parametrization ($w_0, w_a$) in which a time dependent $w(a)$ is Taylor expanded as $w(a) \equiv p / \rho = w_0 + (1-a)w_a$: in this case,  $w_0 = -1.04^{+0.72}_{-0.69}$ and $w_a < 1.32$ at 95 $\%$ C.L. when using Planck + WP + BAO \cite{Planck_params}. As expected, adding $H_0$ data moves these values slightly further  away from $(-1,0)$.  Effectively, $w_a$ tells us how rapidly $\Omega_{de}$ changes from zero to $68 \%$.

Whether Dark Energy was effectively zero or not at early times, can be tested, complementary, using Early Dark Energy \cite{CWP}. Assuming a constant early dark energy $\Omega_e$ at all times from decoupling \cite{doran_robbers_2006} down to when, at recent epochs, a $\Lambda$CDM is restored, provides tight bounds: $\Omega_e \lsim 0.009$ at $95 \% $C.L. for \Planck\ + WP + HighL. Previous bounds, using different datasets, had been found in \cite{calabrese_etal_2011,
reichardt_etal_2011, pettorino_etal_2013}. As shown in \cite{pettorino_etal_2013} such constraints on a constant $\Omega_e$ do not depend on how rapid the transition is from $\Omega_e$ to the present value: it's enough to have an EDE parametrization that depends on $\Omega_e$ only (and not on $w_0$).
On the other hand \cite{pettorino_etal_2013} also showed that such bounds strongly depend on the redshift $z_e$ at which early dark energy becomes non negligible: constraints are substantially weaker if Dark Energy becomes non-negligible only after decoupling. 

In this paper we want to extend the investigation carried out in \cite{Planck_params} to models of dynamical dark energy in which the gravitational interaction between dark matter particles is modified with respect to standard General Relativity. In modified gravity theories, one often has to deal with at least one extra degree of freedom that can be associated to a scalar field, that can be seen as the mediator of a fifth force in addition to standard gravitational interactions. This happens, for example, in scalar-tensor theories (including F(R) cosmologies), massive gravity and all coupled dark energy models, both when matter is involved \cite{amendola_2000, pettorino_baccigalupi_2008} or when neutrino evolution is affected \cite{Fardon:2003eh, Afshordi:2005ym, Amendola_etal_2008, Wetterich:2007kr, Wintergerst:2009fh, Pettorino:2010bv}. Interactions and fifth forces are therefore a common characteristics of many modified gravity models,
the difference being whether the interaction is universal (i.e. it affects all species with the same coupling, as in scalar-tensor theories) or is different for each species (as in coupled dark energy or growing neutrino models). It is therefore interesting to understand the effect of such interactions on the CMB and how large they can be when compared to data. 

In the following we consider the case of coupled dark energy \cite{amendola_2000}, in which dark matter particles feel an interaction mediated by the dark energy scalar field. In this framework, baryons are not affected and still feel standard gravity, while dark matter typically feel a fifth force which is $\beta^2$ times stronger than gravity.
Such an interaction introduces effectively a coupling $\beta$ between the evolution of
the dark energy scalar field and dark matter particles. When seen in the Jordan frame, a coupling between matter and dark
energy can be reformulated in terms of scalar-tensor theories (or $f(R)$  models) \cite{Wetterich:1994bg, pettorino_baccigalupi_2008}.
This is exactly true when the contribution of baryons is neglected.  Alternatively, in the Jordan frame, scalar-tensor theories ($f(R)$ models) require
some sort of screening mechanism (like chameleon \cite{khoury_etal_2004,
Hui:2009kc, 2012arXiv1204.3906U, 2011arXiv1102.5278D} or symmetrons
\cite{2011PhRvD..84j3521H})  that protects the dark energy scalar field and its
mass within high density regions, so that local solar system constraints are
satisfied. To avoid this problem, in the Einstein frame it is instead common use to neglect a coupling to baryon and consider only dark energy - dark matter interactions.

The coupling affects the dynamics of the gravitational potential (and therefore the Late Integrated Sachs-Wolfe effect), the shape and amplitude of perturbation growth, as illustrated in detail in \cite{amendola_etal_2012}. Moreover, the coupling is degenerate with the amount of cold dark
matter $\Omega_{c}$, the spectral index $n$, the Hubble parameter $H(z)$ (see
\cite{amendola_etal_2012} for a review) and can therefore depend very much on the estimates by \Planck\ as well as from the combination of \Planck\ data with other astrophysical datasets that can break these degeneracy.

This paper is organized as follows. In Section II we recall the main features of
coupled dark energy (CDE) cosmologies. In section III we describe the methods used, both with regard to
the implementation of the numerical code and the data used for this paper. In
Section IV we illustrate our results and in Section V we derive our
conclusions.

 \section{Coupled Dark Energy} \label{cde}

Cosmologies in which an interaction is present between dark energy and
dark matter \cite{Kodama:1985bj, amendola_2000, Amendola:2003wa, amendola_etal_2003,
amendola_quercellini_2003, Mangano:2002gg, pettorino_baccigalupi_2008} have to
be seen within the framework of modified gravity, since effectively the
gravitational interaction acting among dark matter particles is modified with respect
to standard General Relativity. Many papers have investigated in details such cosmologies, including spherical collapse (\cite{Wintergerst:2010ui, Mainini:2006zj}
and references therein), higher-order expansions with the time renormalizazion group \cite{Saracco:2009df},
$N$-body simulations \cite{Baldi:2010td,Baldi_etal_2010,Baldi:2010vv},
effects on supernovae, CMB and cross-correlation of CMB and LSS
\cite{amendola_etal_2003,amendola_quercellini_2003, Bean:2008ac,lavacca2009,
Kristiansen:2009yx,Mainini:2010ng,DeBernardis:2011iw,Xia:2009zzb,2013arXiv1305.3106P}
together with Fisher matrix forecasts analysis combining power spectrum and Baryonic Acoustic Oscillations 
measurements as expected by the Euclid satellite \cite{euclidredbook, 2012arXiv1206.1225A} \footnote{http://www.euclid-ec.org/}   and CMB as expected from Planck
\cite{amendola_etal_2012}. The most updated bounds so far were provided in \cite{pettorino_etal_2012} who found $\beta < 0.063$ at $68\%$ confidence level when combining WMAP7+SPT data \cite{Komatsu2011, k11} and first pointed out the impact the a difference in the measurement of $H_0$ between CMB and astrophysical datasets can give in the estimate of the coupling. When adding constraints on the Hubble constant from \cite{riess_etal_2011} a small likelihood peak around $\beta = 0.041$ was found, still compatible with zero at one $\sigma$. The robustness of these constraints has also been tested against a number of tests that investigate the degeneracy with other parameters such as curvature, the relativistic number of degrees of freedom $N_{eff}$, the amplitude rescaled factor of the lensing power spectrum $A_L$ and, most of all, massive neutrinos \cite{lavacca2009, Kristiansen:2009yx, pettorino_etal_2012}. 

We here recall the main equations for coupled dark energy, in order to define the parameters,
and refer to \cite{Amendola:2003wa, pettorino_baccigalupi_2008} for a detailed description of all equations involved. Effects on the CMB have recently been reviewed in \cite{amendola_etal_2012, pettorino_etal_2012}.

\begin{figure}
\includegraphics[scale=0.6]{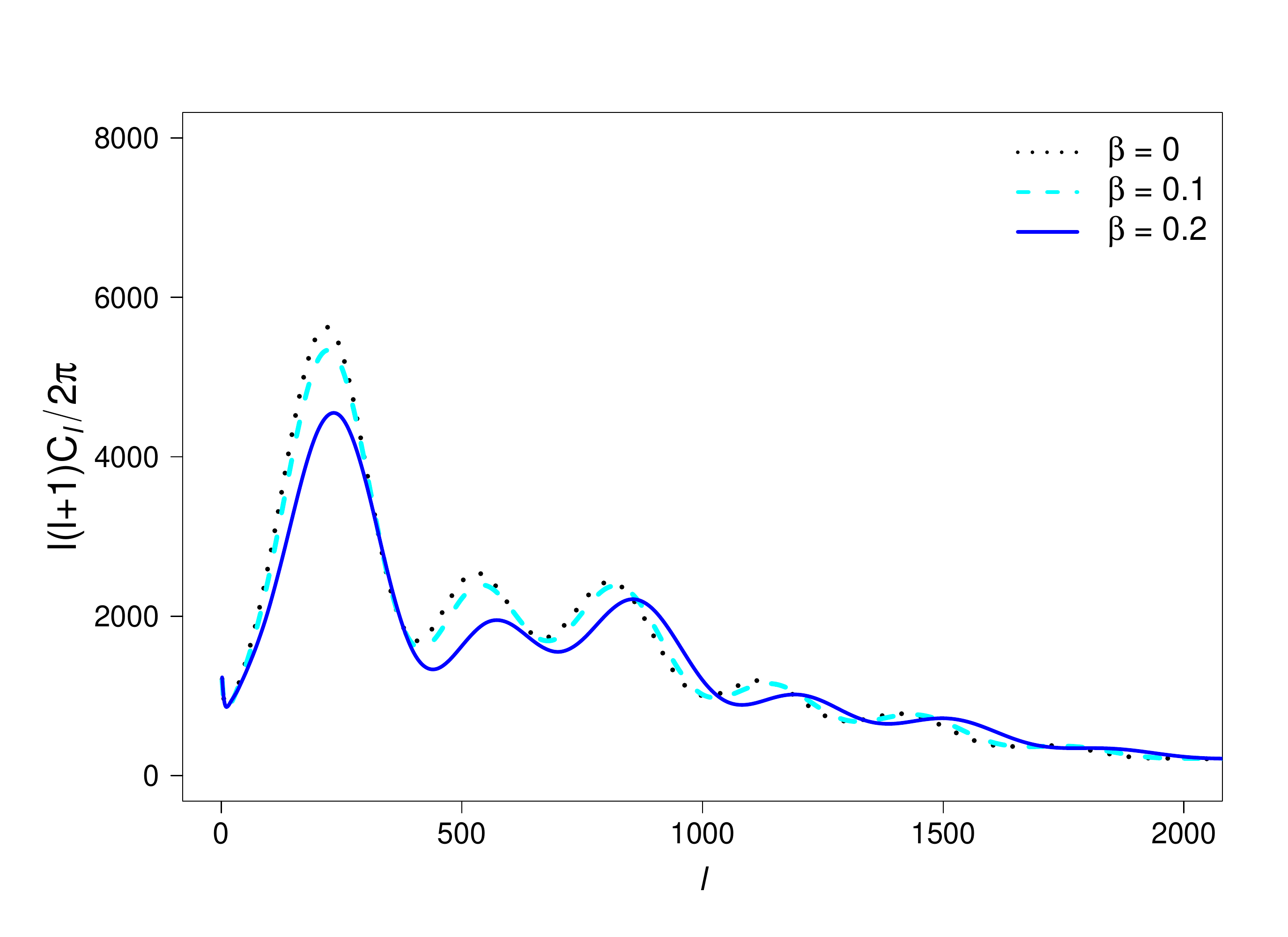}
\caption{\label{fig:CMB-for-three}CMB temperature spectra for
three values of $\beta$ (in agreement with \cite{pettorino_etal_2012}, inserted here for reference).}
\end{figure}

Coupled dark energy cosmologies considered here are described by the
lagrangian:  
 \be \label{L_phi} {\cal L} =
 -\frac{1}{2}\partial^\mu \phi \partial_\mu \phi - U(\phi) -
 m(\phi)\bar{\psi}\psi + {\cal L}_{\rm kin}[\psi] \,, \ee in which the mass of
 matter fields $\psi$ is a function of the scalar field $\phi
 $ and can be related to the coupling $\beta$ as illustrated below. 
Conservation equations for the energy densities of each species read in general as: 
 \bea
 \label{rho_conserv_eq_phi} \rho_{\phi}' &=& -3 {\cal{H}} \rho_{\phi} (1 + w_\phi) - Q \,\,\,\, , \\
 \label{rho_conserv_eq_c} \rho_{c}' &=& -3 {\cal{H}} \rho_{c} + Q \pp
 \nonumber \eea where Q is a generic function. Here we have treated each component as a fluid with
 ${T^\nu}_{(\alpha)\mu} = (\rho_\alpha + p_\alpha) u_\mu u^\nu + p_\alpha
 \delta^\nu_\mu$, where $u_\mu = (-a, 0, 0, 0)$ is the fluid 4-velocity and
 $w_\alpha \equiv p_\alpha/\rho_\alpha$ is the equation of state. Primes denote derivative with respect to conformal time $\tau$.  The class of models considered here corresponds to the choice: 
\be Q = - {\beta} \rho_c \phi' \, , \ee 
the simplest one that can be embedded in a Lagrangian, with an exponential dependence of the mass of dark matter particles on the dark energy scalar field and a constant coupling $\beta$ \cite{amendola_2000, pettorino_baccigalupi_2008}. 
We express the dark energy scalar field in units of the reduced Planck mass $M = (8 \pi G_N)^{-1/2}$. The source Q that appears in the conservation equations (and in the Bianchi identities for these theories) is related to the mass dependence appearing in the Lagrangian: \be Q_{(\phi) \mu} = \frac{\partial
   \ln{m(\phi)}}{\partial \phi} \rho_c \, \partial_\mu \phi . \ee  Equivalently, the scalar field evolves according to the Klein-Gordon equation, which now includes an extra term that depends on CDM energy density:
 \be \label{kg} \phi'' + 2{\cal H} \phi' + a^2 \frac{dV}{d \phi} = a^2 \beta
 \rho_{c} \,\, . \ee
 Throughout this paper we choose an inverse power law potential defined as:
 \be \label{potential}
  V = V_0 \phi^{-\sigma}
 \ee
 with $\sigma$ and $V_0$ constants. The amplitude $V_0$ is fixed thanks to an iterative routine \cite{amendola_etal_2012}
At the level of perturbations, as well as in N-Body simulations, this corresponds to a fifth force that acts among dark matter particles with an effective gravitation constant $G_{eff}$ related to the the standard one $G$ by:
\begin{equation}
G_{eff} = G(1+\beta^2)
\end{equation}
stronger than standard gravity by a factor $\beta^2$.
As discussed in \cite{amendola_etal_2012} the coupling shifts the position
of the acoustic peaks to larger $\ell$'s due to the increase in the
distance to the last scattering surface (this is sometimes called projection effect,
\citep{Pettorino:2008ez} and references therein); furthermore, it reduces the
ratio of baryons to dark matter at decoupling with respect to its
present value, since coupled dark matter dilute faster than in an
uncoupled model. Both effects are clearly visible in Fig.(\ref{fig:CMB-for-three})
for various values of $\beta$ (see also \cite{pettorino_etal_2012}).

Theoretical CMB have been produced implementing the code IDEA
(Interacting Dark Energy Anisotropies) \cite{amendola_etal_2012, pettorino_etal_2012, pettorino_etal_2013} in CAMB \cite{Lewis:1999bs}: these modifications are
able to include dynamical dark energy, Early Dark Energy parameterizations (not
included in this analysis) as well as  interacting dark energy models.
In order to include the coupling, both background and linear perturbations have
been modified following Refs. \cite{Amendola1999, Pettorino:2008ez}.
The output has been compared to an independent
code \cite{amendola_quercellini_2003} that is built on CMBFAST and
the agreement was better than 1$\%$. The difficulty in the implementation relies
on the fact that the initial conditions cannot be obtained analytically as in
simple dark energy parameterizations (early dark energy or $(w_0, w_a)$):
instead, they must be found by trial and error, through an iterative routine
that finds the initial conditions required to get the desired present values of
the cosmological parameters.

We have then performed a Monte Carlo analysis integrating IDEA within COSMOMC
\cite{cosmomc_lewis_bridle_2002} comparing our theoretical predictions with the
data presented in the next section. 

\section{Comparison with observed data}
Constraints on $\beta$ can be obtained if we combine \Planck\ data with astrophysical measurements that break the degeneracy in the distance to last scattering.
As described in \cite{Planck_params} there is some tension in the estimate of the (derived) parameter $H_0$ (the value of the Hubble parameter at present) between CMB data and astrophysical datasets. For a detailed discussion on astrophysical datasets and possible sources of errors we remand to \cite{Planck_params}. In particular, \Planck\ data are more in agreement with BAO than with HST data, when a $\Lambda$CDM cosmology is assumed. In the absence of a known source of this slight discrepancy, we decide here to combine \Planck\ separately either with BAO or with HST data. We do not combine all three datasets and we consider \Planck\ + BAO as the choice in which we can be more confident at present (a sort of conservative choice); we still  evaluate with \Planck\ + HST the impact that HST results would have on our results and in doing so we use \cite{riess_etal_2011}, based on HST observations of Cepheid variables in the host galaxies of eight SNae Ia. This gives a best estimate of $H_0 = (73.8 \pm 2.4)$ km s$^{-1}$ Mpc$^{-1}$ at $1 \sigma$.
Moreover, in including \Planck\ data, we consider two possibilities: 
\begin{enumerate}
\item Planck WP: here we use TT data from Planck plus WMAP low-l polarization;
\item Planck WP + HighL: in this case we also add data from ACT \cite{das_etal_2013} and SPT from \cite{reichardt_etal_2012}, as done in \cite{Planck_params}.
\end{enumerate}
Combining \Planck\ with high-$l$ probes adds information from small scales; these scales are more affected by foregrounds and can be determined by high-$l$ probes with higher precision. The analysis in $\Lambda$CDM provided in \cite{Planck_params} seems to guarantee that foregrounds at small scales are properly accounted also in $\Planck$ data alone and therefore makes us more confident especially when analyzing extensions of the $\Lambda$CDM model. 

The baseline set of parameters includes $\Theta = {\Omega_b h^2, \Omega_{c} h^2, \theta_s, log{\cal{A}}, n_s, \tau}$. These parameters depend on the fractional abundances of the various species, as well as on the amplitude and shape of the primordial power spectrum, and the reionization optical depth; since we  impose spatial flatness, the present dark energy density  $\Omega_{de}$ becomes a derived parameter; 
in addition, coupled dark energy involves two more parameters: $\beta$ and
$\sigma$. Again, $\beta$ represents the coupling between dark matter particles while
$\sigma$ is the parameter in the scalar field potential  (\ref{potential}) that drives the long
range interaction. As illustrated in \cite{pettorino_etal_2012} bounds on $\beta$ do not depend on the value of $\sigma$, which can in turn be written in terms of $w_0$ via the expression $w = -2/(\sigma+2)$; therefore, $\sigma$ can be safely limited to a range in $w_0$ which is still reasonably within observations (w < -0.8). In this sense, we recall that this formulation of coupled dark energy models does not reduce exactly to a $\Lambda$CDM when $\beta = 0$, but rather to a quintessence scalar field in a very flat potential (but not exactly a $\Lambda$CDM).  
The Helium abundance $Y_{He}$ is derived following BBN consistency (see \cite{k11} for details).
As done in \cite{Planck_params}, we assume a minimal-mass normal hierarchy for the neutrino masses, as a single massive eigenstate with $m_\nu =  0.06$ eV. 
We note however that we expect dynamical dark energy (including the coupling) to be partially degenerate with massive neutrinos, as they both contribute to tilt the power spectrum, as it was pointed out in \cite{lavacca2009, amendola_etal_2012}.

\section{Results} \label{sec:results}
\begin{figure*}
 \centering
 \includegraphics[width=18.cm]{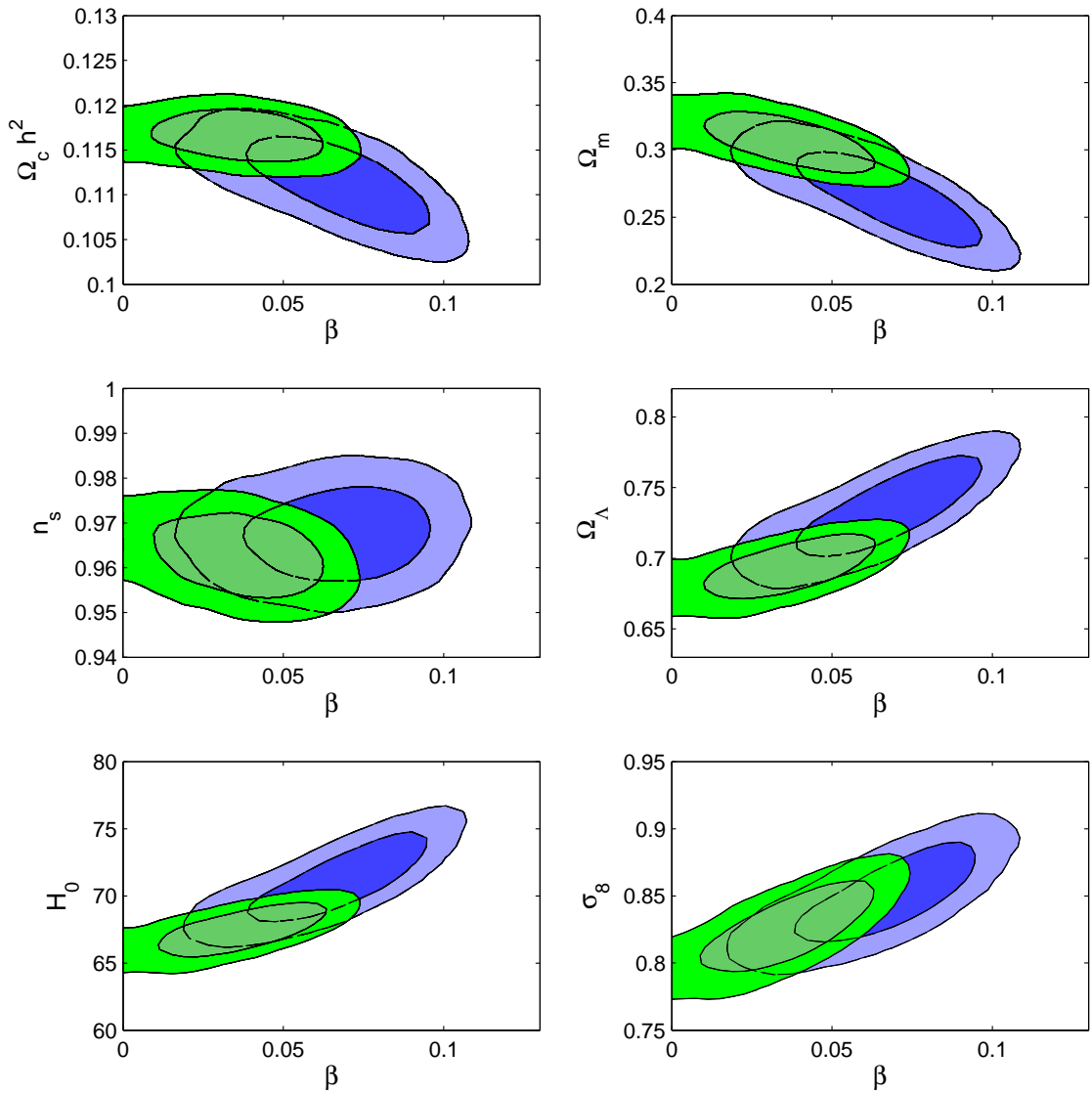}
      \caption{\small Confidence contours for the cosmological parameters for coupled quintessence models. We compare runs PlanckWP + BAO (green) and PlanckWP + HST (blue). 1 $\sigma$ and 2 $\sigma$ contours are shown.}
 \label{fig:like_cont_baseline_cq1_cq2}
 \end{figure*}

Our results from different runs are illustrated in Tab.\ref{tab:runs}. As we can see from the first two columns, the conservative case \Planck\ WP + BAO data has a likelihood peak around a mean value of the coupling $\beta \sim 0.036$, different from zero at roughly $2.2 \sigma$. When adding HighL data, the bound on $\beta$ is roughly the same. This goes along the line pointed out in \cite{Planck_params} and can be seen as confirmation that \Planck\ bounds are stable with respect to foregrounds parameters, whose knowledge (especially of the thermal SZ effect) is expected to be better determined in the HighL data than in \Planck. With respect to a $\Lambda$CDM best fit model, the value of $H_0$ and $\Omega_c h^2$ are not much affected (note that for each parameter we write mean values and not best fits): $H_0 = 67.27 \pm 1.25$ instead of the $\Lambda$CDM one \cite{Planck_params} $H_0 = 67.80 \pm 0.77$ and $\Omega_c h^2 = 0.1169 \pm 0.0020$ instead of $\Omega_c h^2 = 0.1187 \pm 0.0017$.
When we break the degeneracy with HST data the preference for a non zero coupling increases as expected \cite{pettorino_etal_2012}, with a value around $\beta \sim 0.066$ (different from zero at roughly 3.6$\sigma$).
This peak comes mostly from a slight tension between the Hubble parameter HST result ($H_0 =  73.8 \pm 2.4 $ km s$^{-1}$ Mpc$^{-1}$)
and the best fit for $\beta = 0$; it's interesting to note that it's already marginally present in combination with BAO (at about 2.2$\sigma$). We recall, however, that even for $\beta = 0$ we are not in an exact $\Lambda$CDM since in our model w is close, but not exactly equal, to -1.
 \begin{center}
\begin{table}
\begin{tabular}{lllll||}
\textbf{Mean values} & \textbf{for coupled quintessence} & \\
\hline
\\
\begin{minipage}{30pt}
\flushleft
Parameter 
\\
\end{minipage} & 
\begin{minipage}{90pt}
\flushleft
 $\mathbf{PlanckWP}$ \\ $\mathbf{+ BAO}$
\end{minipage} &  
\begin{minipage}{120pt}
\flushleft
 $\mathbf{PlanckWP}$ \\ $\mathbf{+ HighL + BAO}$
\end{minipage} &
\begin{minipage}{100pt}
\flushleft
$\mathbf{PlanckWP}$\\ $\mathbf{+ HST}$
\end{minipage} &  
\begin{minipage}{120pt}
\flushleft
 $\mathbf{PlanckWP}$ \\ $\mathbf{+ HighL + HST}$
\end{minipage}   
\\
\\
\hline
$  \mathbf{\Omega_b h^2} $ & $0.02204 \pm 0.00028$           & 	$0.0221 \pm 0.000269$				& $0.0220 \pm 0.00029$ 	&		$0.0221 \pm 0.000281$	\\
$ \mathbf{\Omega_{c} h^2}  $ &$0.1165 \pm 0.0019$              &	 $0.1169 \pm 0.00197$ 				& $0.1114 \pm 0.00332$	&		$0.1121 \pm 0.00338$ \\
$ \mathbf{\theta_s}$ & $1.0415 \pm 0.000579$                    &		 $1.0415 \pm 0.000576 $ 			& $1.0418 \pm 0.000595$&			$1.0418 \pm 0.000611$ \\
 $\mathbf{ \tau}$ & $0.09037 \pm 0.0132 $                             & 		$0.0904 \pm 0.01267$   				& $0.0913 \pm 0.0135$	&		$0.0936 \pm 0.0126$ \\
  $\mathbf{ n_s}$ & $0.9629 \pm 0.0062 $                              &		 $0.9603 \pm 0.00583$ 				& $0.9677 \pm 0.00673$	&		$0.9655 \pm 0.00678$ \\
   $\mathbf{\beta}$ &  $0.0364 \pm 0.01626$                          &		 $0.0346 \pm 0.0155$ 				& $0.0660 \pm 0.0182$	&		$0.0611 \pm 0.0188$\\
         $\mathbf{\beta}$ &  $0.03132^{+0.0360}_{-0.0266}$       & 	$ 0.0146^{+0.0494}_{-0.0103}$  		& $0.0564^{+0.0409}_{-0.0292}$	&	$0.0708^{+0.0235}_{-0.0514}$ \\
    $\mathbf{ \sigma}$ & $0.2895 \pm 0.1052$                           &	 $0.2837 \pm 0.105$ 				& $0.2932 \pm 0.1055$	&	$0.2681 \pm 0.0996$	\\
   $ \mathbf{\Omega_{de}}$ & $0.6935 \pm 0.0141$                &	 $0.6910^\pm 0.0144 $ 				& $0.7339 \pm 0.0219$	&	$0.7295 \pm 0.0223$ \\
   $ \mathbf{H_0}$ & $67.437 \pm 1.250$                             & 		$ 67.267 \pm 1.247 $   				& $71.123 \pm 2.109$ 	&	$70.737 \pm 2.093$ \\
\hline
\end{tabular}
\caption{For each parameter we report the mean ($\pm$ standard deviation). For $\beta$ we also write the value of the best fit with the two tail errors at $95\%$.
}
 \label{tab:runs}
\end{table}
\par\end{center}
\normalsize
The 2D confidence contours for Planck WP + BAO and Planck WP + HST are plotted in Fig.\ref{fig:like_cont_baseline_cq1_cq2}, where we show a selection of the most interesting likelihood contours vs the coupling $\beta$.  The tension with astrophysical experiments is compensated by an increase in the value of $H_0 \sim 71$ with a corresponding mild decrease of $\Omega_{c} h^2$ and $\Omega_M$, the latter being still compatible with the values estimated in a $\Lambda$CDM scenario. 

 \normalsize

\begin{figure*}
 \centering
 \includegraphics[width=18.cm]{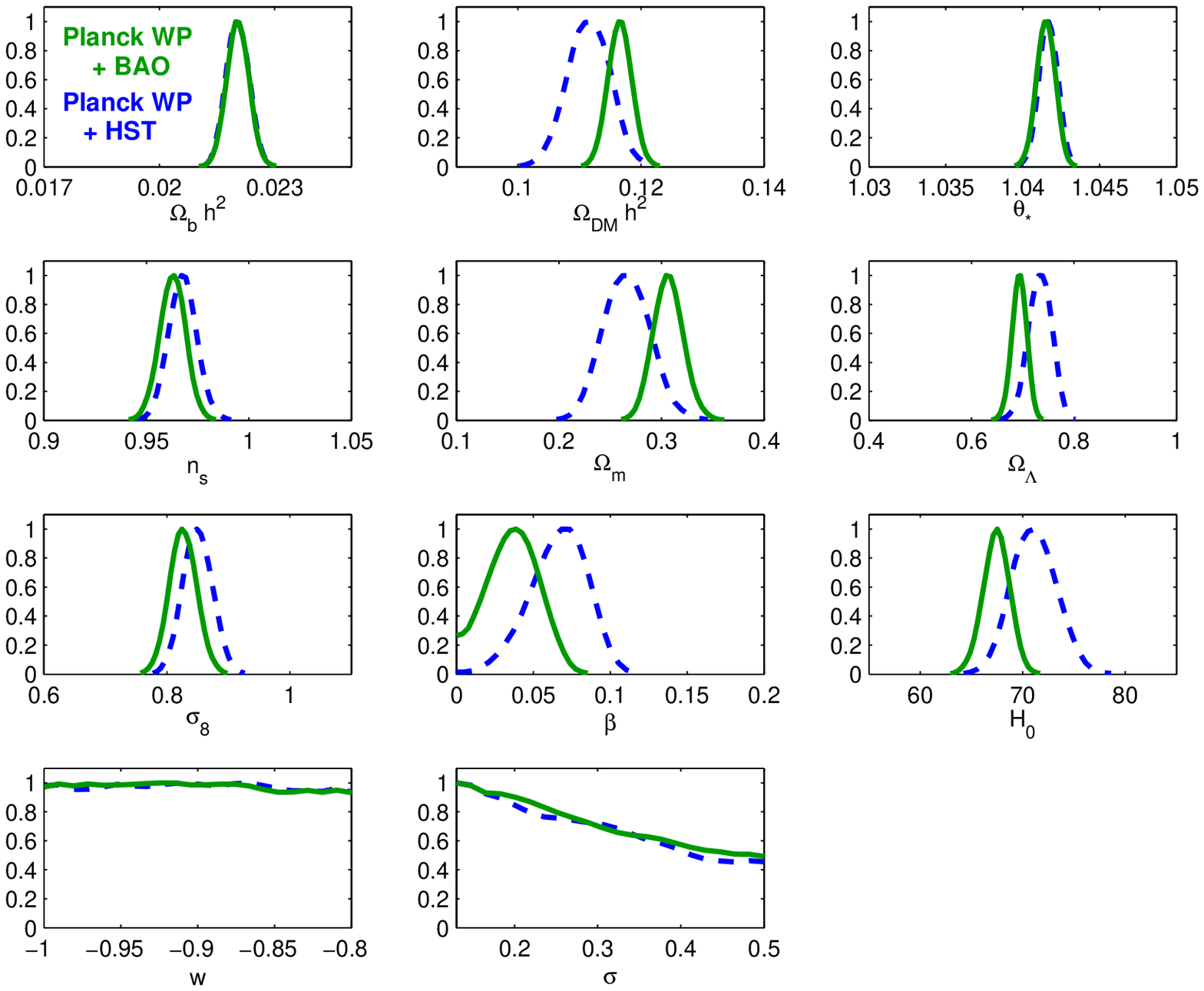}
      \caption{\small 1D likelihoods for the cosmological parameters for coupled quintessence models. We compare runs from Planck WP + BAO (green) and Planck WP + HST (blue). As expected, in coupled quintessence models, no dependence is seen on the value of $\sigma$ and $w$, which are arbitrary and related to each other, as illustrated in the text.}
 \label{fig:like_1D}
 \end{figure*}

In Fig.\ref{fig:like_1D} we also show the corresponding 1D likelihood contours. As expected, there is no determination of $\sigma$, since $\sigma$ only affects late time cosmology. The value of $w$ is arbitrary and approximately related to
$\sigma$ via the expression: $w = - 2 / (\sigma + 2)$; the interval chosen for
$\sigma$ (small enough to get
reasonable speed for the runs) is such that $w$ still assumes reasonable values, at least smaller than
-0.8.
 \normalsize

\begin{figure*}
 \centering
 \includegraphics[width=10.cm]{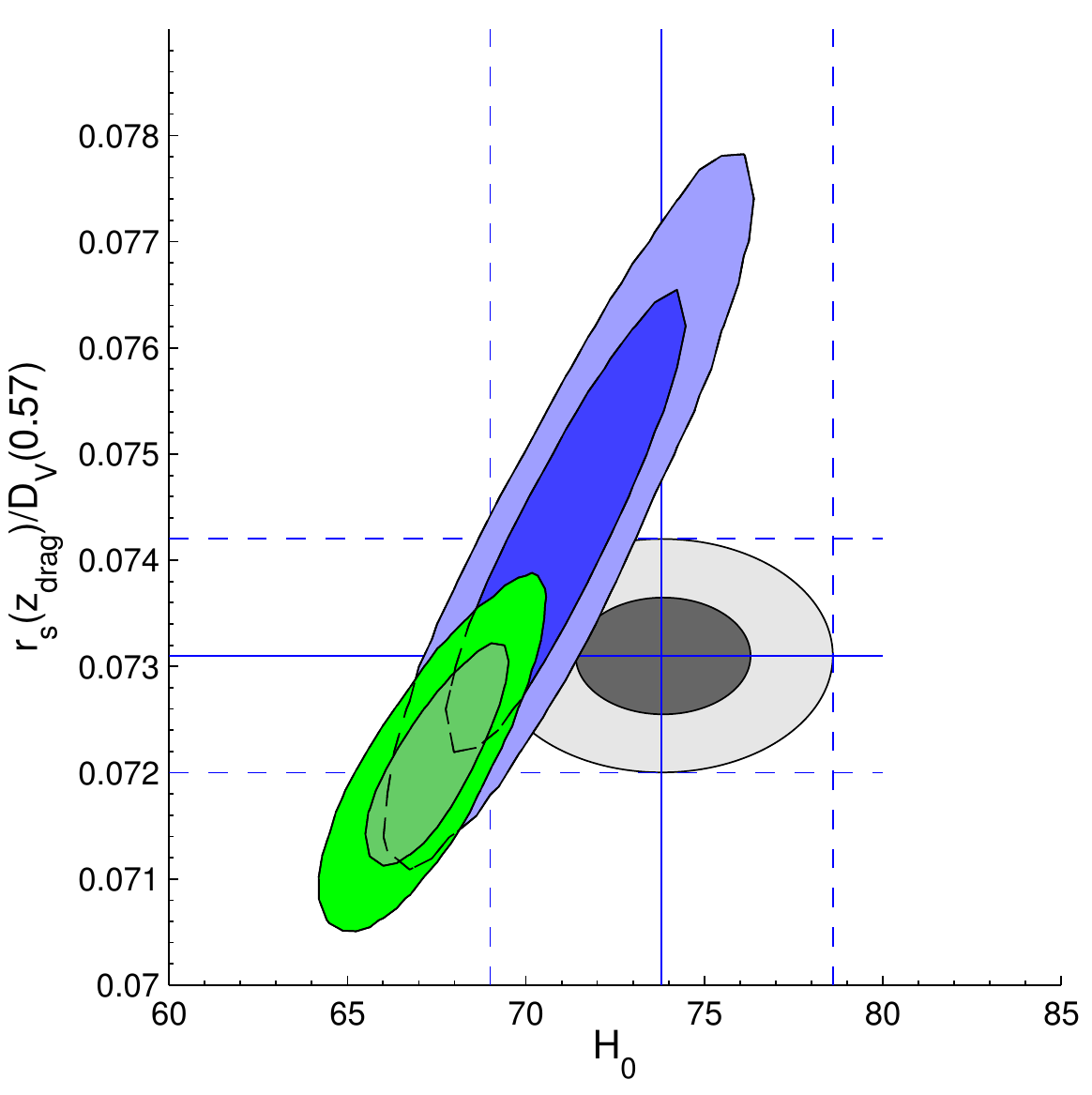}
      \caption{\small This figure (based on Fig.3 of \cite{hou_etal_2012}) investigates the consistency between CMB and BAO$_{BOSS}$ and $H_0$ datasets for a coupled dark energy model. We compare runs \Planck\ WP + BAO (green) and \Planck\ WP + HST (blue). 1$\sigma$ and 2$\sigma$ contours are shown. The grey ellipses correspond to the 1$\sigma$ and 2$\sigma$ region ellipses drawn in Fig.3 of \cite{hou_etal_2012}, which refer to the combination $BAO_{BOSS}$ + $H_0$ (with $H_0$ from \cite{riess_etal_2011}). The horizontal (vertical) solid and dashed lines mark the central value and 1$\sigma$ region.}
 \label{fig:drag}
 \end{figure*}

Finally, in Fig.\ref{fig:drag} we plot the analogous of Fig.3 in \cite{hou_etal_2012} for coupled dark energy models. The figure shows the parameter space $H_0$ vs $r_s/D_V (z_{drag} = 0.57)$. The latter is the characteristic BAO parameter at the redshift reported by BOSS, where $r_{s}$ is the comoving sound horizon at the baryon dragging epoch (when baryons became dynamically decoupled from the photons) and $D_V(z)$ is a combination of the angular-diameter distance $D_A(z)$ and the Hubble parameter $H(z)$:
\begin{equation}  
D_V(z) = \left[ (1+z)^2 {D_A}^2 \frac{c z}{H(z)}\right]^{1/3} \,\,\, .
\end{equation}
While the green and blue contours refer to CMB likelihoods from \Planck\ WP + BAO and \Planck\ WP + HST data respectively, the grey ellipses show the ellipses from BOSS+$H_0$ (using \cite{riess_etal_2011} for the latter), as reported in \cite{hou_etal_2012}. The ellipses partly overlap and are compatible at about two $\sigma$. We prefer however not to combine all three probes (CMB, BAO, $H_0$) and wait for further clarifications on these datasets, as discussed in \cite{Planck_params}. We just note that a larger coupling corresponds to larger values of $H_0$, as seen by the clear degeneracy plotted in fig.\ref{fig:like_cont_baseline_cq1_cq2}.

 \normalsize

In order to have a feeling on how the coupling $\beta$ is related to other common measurements of gravitational interactions, constraints on $\beta$ can be converted into a constraint on the Brans-Dicke coupling parameter $\omega_{BD}$ as shown in \cite{amendola_etal_2012} using:
\begin{equation}
3 + 2 \omega_{BD} = \frac{1}{2 \beta^2}
\end{equation}
Therefore a value of $\beta \sim 0.036 (0.066)$ would correspond to $\omega_{BD} \sim 191 (56)$ and to a post-Newtonian parameter $\gamma_{PPN} = 1-2/(2\omega_{BD}+3) \sim 0.99 (0.98)$. These values have to be seen as complementary to the small-scale limits set by local gravity tests on Yukawa corrections \cite{2007PhRvL..98b1101K, Nakamura:2010zzi} and refer to a dark matter - dark matter interaction on cosmological scales (baryons are assumed to follow general relativity, as explained above).

\section{Conclusions} \label{conclusions}
We have considered the possibility that the evolution of dark matter and dark energy might be
connected by a constant coupling, of the type illustrated in
\cite{amendola_2000, pettorino_baccigalupi_2008}. This effectively introduces a fifth force that modifies the gravitational attraction between dark matter particles. 
We have used current CMB data
from Planck to constrain the coupling parameter $\beta$. This parameter measures the amount by which gravitational interaction between dark matter particles is modified. Constraints on $\beta$ are complementary to the small-scale limits set by local gravity tests on Yukawa corrections \cite{2007PhRvL..98b1101K, Nakamura:2010zzi}.
Due to the degeneracy with the distance to last scattering, we combine CMB data with different astrophysical datasets (BAO or HST).
 We find that a small preference for non-zero coupling $\beta$, less or more significant depending on the astrophysical dataset used. In particular we find $\beta = 0.036 \pm 0.016$  at $68\%$  C.L. for Planck WP + BAO and $\beta = 0.066 \pm 0.018$ for Planck WP + HST data. These values are in less or more tension with zero at roughly 2.2$\sigma$ or 3.6$\sigma$ respectively. It is interesting to notice that a small preference for a non-zero coupling is present also when combining \Planck\ with BAO, whose astrophysical geometrical measurements seems to be more reliable \cite{Planck_params}. Given the number of possible systematics which may affect datasets our attitude is to be conservative: we do not find this preference strong enough to claim a deviation from a $\Lambda$CDM. Our aim here is mainly to show that CMB data, though compatible with a $\Lambda$CDM, still contain significant information that does not exclude the presence of dynamical Dark Energy models and fifth forces. After completion of this paper, the article \cite{2013arXiv1304.7119S} was published on the ArXiv, finding similar results for a different set of coupled dark energy models, based on a coupling inserted through conservation equations rather than at the level of the Lagrangian (\ref{L_phi}).

\begin{acknowledgments}
V.P. is supported by the Marie Curie IEF, Project DEMO - Dark Energy Models and Observations. The initial idea and motivation for this paper was discussed with L.Amendola, to whom I am grateful for precious comments. I also thank Claudia Quercellini for useful collaboration on the original IDEA code.
\end{acknowledgments}
\bibliography{pettorino_bibliography}

\end{document}